# Pattern Formation in Reaction-Diffusion System on Membrane with Mechanochemical Feedback

*Naoki Tamemoto and Hiroshi Noguchi\**

Institute for Solid State Physics, University of Tokyo, Kashiwa, Chiba 277-8581, Japan

Correspondence and requests for materials should be addressed to H.N. (noguchi@issp.u-tokyo.ac.jp).

**ABSTRACT**

Shapes of biological membranes are dynamically regulated in living cells. Although membrane shape deformation by proteins at thermal equilibrium has been extensively studied, nonequilibrium dynamics have been much less explored. Recently, chemical reaction propagation has been experimentally observed in plasma membranes. Thus, it is important to understand how the reaction-diffusion dynamics are modified on deformable curved membranes. Here, we investigated nonequilibrium pattern formation on vesicles induced by mechanochemical feedback between membrane deformation and chemical reactions, using dynamically triangulated membrane simulations combined with the Brusselator model. We found that membrane deformation changes stable patterns relative to those that occur on a non-deformable curved surface, as determined by linear stability analysis. We further found that budding and multi-spindle shapes are induced by Turing patterns, and we also observed the transition from oscillation patterns to stable spot patterns. Our results demonstrate the importance of mechanochemical feedback in pattern formation on deforming membranes.

**INTRODUCTION**

Membrane deformation is a fundamental biological process involved in many cellular functions such as vesicular transport[1], cell division[2], and cell motility[3]. To understand these phenomena, the mechanism of membrane deformation by intracellular proteins has been investigated in detail[4–8]. Recently, it has been shown that the deformation of biological membranes is not just a passive phenomenon but also plays physiological roles[8–15]. For example, membrane curvature induces localization of membrane proteins in highly curved domains[9] and phase separation of lipid membranes[10–12]. This clustering can lead to the emergence of lipid rafts, which are believed to play important roles in cell signaling and membrane trafficking[12,13,16]. Membrane binding by curvature-inducing proteins that are involved in vesicular transport is also regulated by membrane curvature and by various proteins[7,17,18]. For example, recruitment of curvature-inducing protein FBP17, involved in endocytosis, onto the membrane is regulated by the local membrane curvature, membrane tension, and endocytic proteins[17–19]. This mechanism is suggested to play important roles in cell polarization[19], endocytosis[20], and cell division[21].

To understand pattern formation on curved surfaces, several types of studies have been conducted[22–32]. One typical approach is to analyze pattern formation at thermal equilibrium based on phase separation[22–26]. This type of study has shown that membrane shapes and domain patterns of equilibrium states are affected by the line tension of domain boundaries, bending rigidity, and local curvatures[22–26]. Such studies have successfully described the experimentally observed patterns of multi-component lipid vesicles. However, studies pertaining to kinetics are limited to the dynamics of relaxation toward an equilibrium state[22–26,31,33].

Most of previously conducted theoretical and numerical studies have examined only the effects of protein binding; however, in living cells, it is known that many proteins typically work in concert to regulate biological functions. Propagation waves in membranes are often observed during cell migration, spreading, growth, or division[34–41]. Such waves and chemical patterns can be reproduced through activator-inhibitor systems of reaction-diffusion models[42]. The reaction-diffusion system was first proposed by Turing to describe the symmetry breaking of morphogenesis[43], and has been applied to curved surfaces such as animal skins and tissues[44–46]. These studies have shown that geometry affects pattern formation and domain localization;[29] however, the conclusions of such studies are limited by the fact that the surface shape is fixed, although the effects of size increase have been investigated[27,28]. Recently, the propagating waves of F-BAR protein and actin growth have been explained by the reaction-diffusion systems of five chemical reactants on a quasi-flat membrane[18]. As large membrane deformations caused by the coupling of curvature and reaction-diffusion systems have not yet been studied[41], the effects of membrane deformation on reaction-diffusion systems have not been elucidated.

In this study, we investigated the coupling effects between membrane deformation and reaction-diffusion systems by simulating vesicle deformation through curvature-inducing proteins and also chemical reactions using a reaction-diffusion model. Our model accounts for the mechanochemical feedback between membrane curvature and protein concentration. We employed a dynamically triangulated surface model to represent the membrane and calculated the curvature energy to solve the membrane deformation dynamics[47–49]. We employed the Brusselator model[50], one of the simplest reaction-diffusion systems, modifying it to include the mechanochemical feedback from membrane curvature. As the dynamics of a non-deformable surface are well understood, we were able to analyze the evident membrane-deformation effects. We describe how this coupling changes the vesicle shape and pattern formation.

**RESULTS**
**Reaction-diffusion model and stability analysis.**

A two-dimensional reaction-diffusion system with two reactants is written as $\tau \frac{\partial u}{\partial t} = D_u \Delta u + f(u,v)$ and $\tau \frac{\partial v}{\partial t} = D_v \Delta v + g(u,v)$, where $\tau$ is a time constant, $D_u$ and $D_v$ are diffusion coefficients of reactants $u$ and $v$, and $\Delta$ is a two-dimensional Laplace–Beltrami operator. In this study, we consider the Brusselator model, which is described by the reaction scheme below:

$$A \to u$$
$$B + u \to v$$
$$2u + v \to 3u$$
$$u \to E.$$

The reaction equations are given by $f(u,v) = A - (B+1)u + u^2 v$ and $g(u,v) = Bu - u^2 v$, where $A$ and $B$ are positive parameters[50].

In the coupling of the reaction-diffusion system with the change in membrane curvature, $u$ represents the local area fraction covered by curvature-inducing binding proteins on the membrane ($u \in [0,1]$), and $v$ is the concentration of a protein to regulate the protein binding. The free energy in relation to curvature is expressed as $F_{cv} = \int f_{cv} dS$, with

$$f_{cv} = (1-u)\frac{\kappa_0}{2}(2H)^2 + u\frac{\kappa_1}{2}(2H - C_0)^2, \quad (1)$$

where $\kappa_0$ and $\kappa_1$ represent the bending rigidity without or with the bound proteins, respectively; $C_0$ is the spontaneous curvature; $S$ is the surface area; and $H$ is the mean curvature, $H = (C_1 + C_2)/2$, where $C_1$ and $C_2$ are two principal curvatures. The corresponding curvature term $A'$ is added to the reaction equation $(u,v)$; thus the reaction-diffusion equations are written as

$$\tau\frac{\partial u}{\partial t} = D_u \Delta u + \frac{A + A'}{k_u} - (B+1)u + k_u u^2 v \text{ and } A' = -G\frac{\partial f_{cv}}{\partial u}, \quad (2)$$

$$\tau \frac{\partial v}{\partial t} = D_v \Delta v + B k_u u - k_u{}^2 u^2 v, \tag{3}$$

where $G$ is the mechanochemical feedback magnitude of the reaction ($G \geq 0$), and $k_u$ is a normalization factor expressed as $k_u u$, used to obtain Turing and oscillation phases at $u \in [0,1]$. To maintain $0 \leq u \leq 1$, $u$ is restricted between the lower and upper bounds: it is set to $u = 0$ or $u = 1$ when the time evolution of Eq. (2) crosses those bounds. The first reaction becomes $A + A' \rightarrow u$, which can be considered to represent the binding of protein $u$ from the solution surrounding the membrane. Thus, the binding of $u$ is enhanced at a membrane curvature $H \simeq C_0/2$, where $\frac{\partial f_{cv}}{\partial u} < 0$ so that $A' > 0$. On the other hand, the time evolution of $v$ is not directly dependent on the local membrane curvature. Note that the mixing-entropy term of the protein concentration is not accounted to reproduce the normal Brusselator dynamics when the membrane shape is fixed. In this study, we use $A = 4.5$, $B = 2.02$, $\eta = \sqrt{D_u/D_v} = 0.1$, and $k_u = 4.52$ for all simulations.

Based on the linear stability analysis around the fixed point, $(u_s, v_s) = \big((A + A')/k_u, B/(A + A')\big)$[51], the conditions for Hopf and Turing bifurcations with a membrane curvature effect $A'$ on a fixed spherical surface are $B > 1 + (A + A')^2$ and $B > (1 + (A + A')\eta)^2$, respectively; and temporal oscillations and spatial patterns appear above these. The membrane curvatures for Hopf and Turing bifurcations are given below, respectively:

$$2(A - \sqrt{B-1}) < GH^2 E_{cv} \text{ and,} \tag{4}$$
$$2(A + (1 - \sqrt{B})/\eta) < GH^2 E_{cv}, \tag{5}$$

where $E_{cv} = \kappa_1 (C_0/H)^2 - 4\kappa_1 C_0/H + 4(\kappa_1 - \kappa_0)$. At $A + A' < 0$, i.e., $2A < GH^2 E_{cv}$, a homogeneous phase is formed, because $u_s < 0$. The phase stability diagram is shown in Fig. 1. This diagram shows that bifurcations occur as the magnitude of the spontaneous curvature $C_0$ and mechanochemical coupling magnitude $G$ increase at $A + A' \geq 0$.

**Pattern formation on membrane.**

The membrane motion is solved by the Langevin dynamics of dynamically triangulated surface model, which formed a triangular network of spherical topology with $N$ vertices, as described previously[47]. In this study, the presence of curvature-inducing proteins is considered in addition to the model as given in Eq. (1). We use $\kappa_0/k_B T = 20$ and $\kappa_1/k_B T = 40$, where $k_B T$ is the thermal energy (see Methods for more details). The results are displayed with the length unit $R = \sqrt{S/4\pi}$, energy unit $\kappa_0$, and time unit $\tau$.

First, we analyzed the pattern formation on the fixed surface of a spherical vesicle at the reduced volume, $V^* = 3V/4\pi R^3 = 1$, where $V$ is the vesicle volume (Figs. 2(a), (b), and (g)). The results are consistent with those of the linear stability analysis (Fig. 2(g)). The effects of thermal fluctuations are discussed in the Supplementary Material. Figs. 2(a) and (b) show typical snapshots. One large circular Turing domain appears at $G\kappa_0/R^2 = 0.061$ and $C_0 R = 8$ (Fig. 2(b)).

In contrast, membrane deformation changes the chemical patterns in deformable vesicles at $V^* = 0.8$ (Figs. 2(c)–(f), and (h)). The oscillation phase is suppressed, and Turing pattern is observed in a wider parameter region (Fig. 2(h)). At high spontaneous curvature $C_0$, budding and spicule shapes are formed, accompanied by Turing patterns (Figs. 2(d) and (e)). These spicule shapes only appear under conditions of Turing pattern formation, while budding can occur in homogeneous membranes. Moreover, budded spheres typically have a high value of $u$ that is homogeneously distributed and form a Turing domain boundary separating two phases with higher or lower value of $u$ at the narrow connective neck, as shown in Fig. 2(f), because of the reduction in diffusion through the neck. Thus, the Turing pattern is modified by the membrane shape deformation. Bud formation is obtained for $C_0 R \geq 3$ at $V^* = 0.8$ (Fig. 2(h)). This is reasonable, as the curvature energy of a spherically shaped bud with a radius $r_b = 2/C_0$, which is fully covered by the curvature-inducing protein ($u = 1$) is minimal. The condition of bud formation is given by $V^* \leq (r_b/R)^3 + (1 - (r_b/R)^2)^{3/2}$, since the volume of the rest of a vesicle of a spherical shape is maximal. In the case of $V^* = 0.8$, the threshold is $R/r_b \geq 2.2$.

For high values of $C_0$, different shapes can be formed depending on the initial shapes, such as the prolate and budded shapes shown in Fig. 2(h). Figure 3 shows another example. Vesicles of three or four spicules are formed from prolate and oblate vesicles, respectively, with $(u, v) \simeq (u_s, v_s)$ (Figs. 3(a) and (b)). As pattern formation progresses, the vesicle shape changes according to the chemical pattern. In order to evaluate the non-uniformity of $u$ and the smoothed local curvature $\tilde{H}$, we calculated separation metrics, $s_u = \sigma_b(u)^2/\sigma_w(u)^2$ and $s_H = \sigma_b(\tilde{H})^2/\sigma_w(\tilde{H})^2$, where $\sigma_b^2$ and $\sigma_w^2$ are the between-class variance and within-class variance, respectively[52] (The curvature smoothing method is described in the Supplementary Material). Each variance is calculated as below:

$$\sigma_b^2 = \rho_0 \rho_1 (\mu_0 - \mu_1)^2 \text{ and,} \tag{6}$$
$$\sigma_w^2 = \rho_0 \sigma_0^2 + \rho_1 \sigma_1^2, \tag{7}$$

where $\rho_i$ is the probability of each class, $\mu_i$ is the class mean value, and $\sigma_i^2$ is the class variance. The threshold value to divide into two classes is determined to maximize the metric value. Therefore, the metrics $s_u$ becomes large when Turing patterns are formed clearly, whereas $s_u$ becomes small when the two phases are gently separated or not separated (i.e., homogeneous patterns). Figures 3(c) and (d) show that $s_u$ increases as the Turing pattern develops, followed by an increase in $s_H$; this sequence is consistent with that depicted by the sequential snapshots and indicates that non-uniformity can be distinguished by calculating the separation metrics. We also calculated the time development of asphericity, $\alpha$, to evaluate vesicle deformation (Fig. 3(e)). Asphericity is the degree of deviation from a spherical shape, calculated as below:

$$\alpha = \frac{(\lambda_1 - \lambda_2)^2 + (\lambda_2 - \lambda_3)^2 + (\lambda_3 - \lambda_1)^2}{2(\lambda_1 + \lambda_2 + \lambda_3)^2}, \qquad (8)$$

where $\lambda_i$ is the eigenvalue of the gyration tensor of the vesicle[47,53,54]. For a sphere, $\alpha = 0$ ($\lambda_1 = \lambda_2 = \lambda_3$), and for the thin-rod limit, $\alpha = 1$ ($\lambda_1 = 1$ and $\lambda_2 = \lambda_3 = 0$). As the vesicle forms three or four spindles, $\alpha$ decreases (Figs. 3(e) and (f)).

To further investigate the effect of coupling between the Brusselator and vesicle deformation, we conducted the simulation with different $C_0$ and $D_u$ values at $G\kappa_0/R^2 = 0.046$ (Fig. 4 and Supplementary Fig. S3). As $C_0$ decreases, the number of domains, $N_d$ and $s_H$ decrease, whereas $\alpha$ increases (Figs. 4(e–g)). In addition, the domain size increases as $D_u$ increases. Therefore, higher $N_d$ and $s_H$ values and a lower $\alpha$ are obtained at a lower $D_u$ (Figs. 4(h–j)). When $N_d > 2$, convex regions are formed in various directions and the vesicle becomes nearly spherical, but when $N_d = 2$, the vesicle becomes prolate in shape, and $\alpha$ increases (Figs. 4(a–d)). Thus, chemical pattern formation affects vesicle deformation and the relation between $N_d$ and the preferred curvature of the domains is important in determining the stable shapes. The results do not significantly differ between simulations that start from prolate or oblate shapes, except under the condition at $C_0 R = 7$ and $D_u = 20$ (Figs. 4(a) and (f)). Under that condition, with starting from a prolate-shaped vesicle, two domains arise at the pole of prolate, and the vesicle shape remains in the prolate shape. However, when the simulation starts from the oblate-shaped vesicle, multiple domains arise at the edge of oblate, and vesicle shape morphs into a multi-spindle shape (Figs. 4(a), (f), and (g)). As well as the effect of chemical pattern formation to the vesicle deformation, vesicle shape can also affect chemical pattern formation.

A comparison of Fig. 2(g) with Fig. 2(h) shows that the region encompassing Turing patterns is enlarged in the phase diagram at $V^* = 0.8$ from $V^* = 1$, as $G$ increases. To investigate this change, we performed simulations at $G\kappa_0/R^2 = 0.077$ and $C_0 R = 10$ with different $V^*$ and $D_u$ (Fig. 5). For $D_u = 10$ or $20$, Turing patterns occur instead of oscillations, whereas for $D_u = 50$, an oscillation occurs at $V^* = 0.95$, and the oscillating patterns transition to the Turing pattern at $V^* = 0.8$ and $0.65$ (Figs. 5(d), (g) and (j)). As shown in Figs. 5(e), (h), and (k), the maximum values of the local curvature $\tilde{H}_{max}$ at $V^* = 0.8$ and $0.65$ eventually increase over time; this does not occur at $V^* = 0.95$. As the local curvature $H$ increases, the position on the phase diagram shifts toward the upper left, as shown in Supplementary Figure S5. Therefore, the transitions from an oscillation pattern to a Turing pattern is induced by a local increase in $H$.

At $V^* = 0.95$, a small domain is generated and stabilized by the local deformation of the vesicle at $D_u = 10$. In contrast, a large domain is temporarily generated at $D_u = 50$, but is not stabilized, since the stable domain size is much larger than the sphere of preferred curvature $C_0/2$; thus the vesicle cannot sufficiently deform (Fig. 6). In addition, the oscillation period for $D_u = 50$ is significantly longer for $V^* = 0.95$ than for $V^* = 0.8$ or for $V^* = 0.65$ (Fig. 5). The oscillation period $\tau_{\mathrm{os}}$ is calculated from the peak of the Fourier spectrum of $s_u$ for the eight independent runs: $\tau_{\mathrm{os}}/\tau = 100$, 11, and 8 at $V^* = 0.95$, 0.8, and 0.65, respectively. This fact and the time evolution of $\widetilde{H}_{\max}$ indicate that membrane deformation is suppressed by the volume restriction for $V^* = 0.95$ (Fig. 5(e)). In contrast, substantial membrane deformation occurs at the reduced volumes of $V^* = 0.8$ and 0.65, which enables frequent generation of domains. Thus, membrane deformation can change both oscillation period and stability.

**DISCUSSION**

In this study, we have examined the coupling effects between a reaction-diffusion system and membrane deformation by simulating membrane deformation using a dynamically triangulated surface model. We adapted the Brusselator model to include mechanochemical feedback between local membrane curvature and the concentration of curvature-inducing proteins. Based on the linear stability analysis of the reaction-diffusion system with a membrane curvature effect on a fixed spherical surface, we have clarified that bifurcation curves depend on the mechanochemical coupling magnitude $G$ and the value of spontaneous curvature of curvature-inducing proteins $C_0$ with respect to the local membrane curvature (Fig. 1). Thus, the stability of both Turing and oscillation dynamics depend on the membrane shape. We have shown that various shapes, such as buds and multi-spindles, depend on $G$, $C_0$, and the diffusion constant $D_u$ (Figs. 2–4). In addition, since the domain formation of curvature-inducing proteins is promoted at regions with high local curvature, the initial shape of the vesicles affects the dynamics of pattern formation (Fig. 4(a)). Therefore, the dynamics of protein pattern formation change the shape of vesicles, while membrane deformation simultaneously affects pattern formation. This feedback loop can drastically alter the chemical reaction patterns from those on non-deformable surfaces (Figs. 2(g) and (h)). A dynamic transition from an oscillating pattern to a Turing pattern is induced by membrane deformation (Figs. 5(g–i)). Such transitions have not been reported in previous studies.

In the context of living cells, many kinds of proteins and other molecules function interdependently on membranes, where the function of one protein is often activated or inhibited by those of others. Membrane deformation brought about by competing forces of protein-induced curvature changes and surface tension changes impelled by actin growth has been studied[4,8,18,19]. By choosing not to consider the dynamics of actin in this study, we demonstrated that various membrane deformations, accompanied by Turing patterns and oscillations, can be produced by one curvature-inducing protein and one or a small number of regulatory proteins without actin interactions.

Here, we analyzed the coupling of a reaction-diffusion system with membrane deformation utilizing the fixed parameters $A$, $B$, and $\eta$, focusing primarily on Turing patterns, and oscillatory conditions to a lesser extent. The experimental results indicate that observed patterns, which include a feedback loop between curvature-inducing proteins and membrane deformation, are not only stable spot patterns, such as those observed during cell polarization[19], but are also propagating waves[18]. Similarly, the reconstituted Min system in liposomes, which regulates bacterial cell division, has been shown to exhibit propagating wave patterns[38–40]. These patterns, which induce oscillating membrane deformation, are also described by reaction-diffusion systems. The system developed in this paper can also be applied to these patterns observed in living systems, by adjusting the parameters. Other chemical reaction models, such as the Oregonator[55], which was developed to model the Belousov−Zhabotinsky reaction, and the F-BAR–actin model[18], are also easily applied. Thus, the present model system is a powerful tool that can be used to study a wide range of chemical reaction systems that are coupled with membrane deformation.

**METHODS**
**Membrane model.**

Membrane contains $N = 4000$ vertices connected by bonds of an average length $a$, with volumes and masses, $m$, excluded. The local curvature energy $f_{\text{cv}}$ in Eq. (1) is discretized using dual lattices. The surface area $S = 0.41a^2(2N - 4) \simeq 3280a^2$ and volume $V$ of a vesicle are kept constant at about 0.01% accuracy by harmonic constraint potentials. Details of the potentials are described in Ref. 47. For the coefficients of area and volume constraint potentials, four times greater values are employed than those in Ref. 47. To produce membrane fluidity, bonds are flipped to the diagonal of two adjacent triangles using the Monte Carlo method. The membrane motion is solved by molecular dynamics (MD) with the Langevin thermostat:

$$m\frac{\partial^2 \boldsymbol{r}_i}{\partial t^2} = -\frac{\partial U}{\partial \boldsymbol{r}_i} - \zeta\frac{\partial \boldsymbol{r}_i}{\partial t} + \boldsymbol{g}_i(t), \tag{9}$$

where $\zeta$ is the friction coefficient, and $\boldsymbol{g}_i(t)$ is Gaussian white noise, which obeys the fluctuation-dissipation theorem. The hydrodynamic interactions are not considered. The time unit in MD is $\tau_{\text{md}} = \zeta a^2/k_B T$ based on diffusion, and $m = \zeta\tau_{\text{md}}$ is used. To allow membrane deformation followed by concentration changes in $u$, $\tau_{\text{md}} = 0.1\tau$ is employed. Equation (9) is numerically integrated by the leapfrog algorithm with time steps $\Delta t_{\text{md}} = 0.001\tau_{\text{md}}$.

**Discretization of reaction-diffusion equations.**

We developed a finite volume method to discretize Eqs. (2) and (3). Since the Kelvin-Stokes theorem holds for curved surfaces, it is straightforwardly applicable, as employed on a flat surface. A vertex-centered finite volume approach is applied to the dual lattices used for the calculation of membrane curvature[47]. The time evolution of $u$ of the *i*-th vertex is discretized using the following forward difference method:

$$u_i(t + \Delta t_{\rm rd}) = u_i(t) + f(u,v)\Delta t_{\rm rd} + D_u \Delta t_{\rm rd} \sum_j (u_j - u_i) \frac{l_{ij}}{r_{ij} S_i}, \qquad (10)$$

where $S_i$ is the vertex area, $l_{ij}$ is the side length between neighboring vertex cells, and $r_{ij}$ is the distance between the neighboring vertices. The effect of curvature on diffusion is included as the variation of side lengths. Similarly, Eq. (3) is discretized. In this study, $\Delta t_{\rm rd} = 0.1 \Delta t_{\rm md}$ is used. The initial concentrations for the simulations are set around the fixed point $(u_{\rm s}, v_{\rm s})$, with small random perturbations. When $u_{\rm s} < 0$ or $u_{\rm s} > 1$, $u = 0$ or $u = 1$ are taken, respectively, as the initial concentration instead.

## AUTHOR INFORMATION

**Corresponding Author**

*E-mail: noguchi@issp.u-tokyo.ac.jp

## ACKNOWLEDGMENTS

We would like to thank H. Kitahata (Chiba Univ.) and H. Kori (Univ. Tokyo) for stimulating discussions. This work was supported by JSPS KAKENHI Grant Number JP17K05607.

**Author Contributions**

N.T. and H.N. designed the research. N.T. performed the computations and analyzed the data. N.T. and H.N. wrote the manuscript.

Competing Interests: The author declares no competing interests.

**FIGURES**

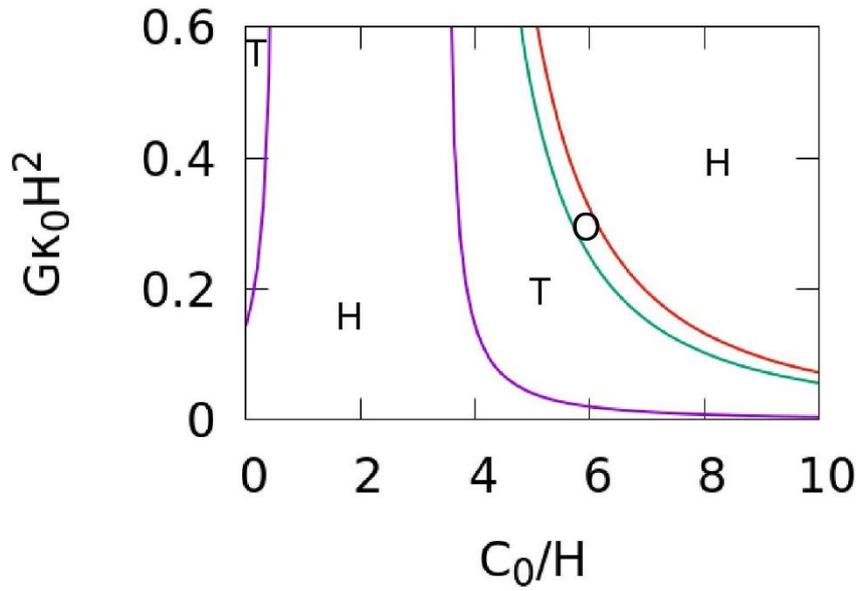

Figure 1. The phase diagram for the Brusselator, modified to include a membrane curvature effect, on a surface of a constant mean curvature $H$ at $A = 4.5$, $B = 2.02$, $\eta = 0.1$, and $\kappa_1/\kappa_0 = 2$. The purple and green lines are the Turing bifurcation curve and the Hopf bifurcation, respectively. These curves separate the regions in which the homogeneous stable patterns (H), stationary Turing patterns (T), or temporal oscillation patterns (O) occur. The red line indicates $A + A' = 0$.

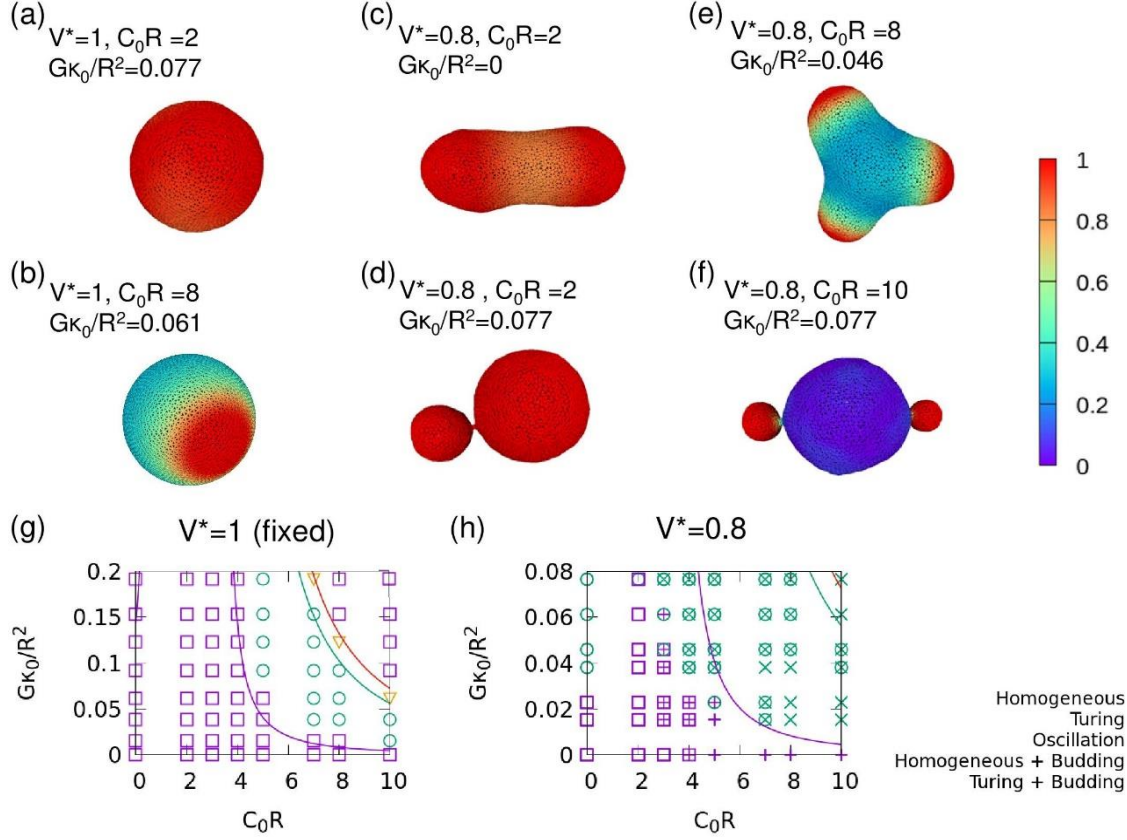

Figure 2. (a–f) Snapshots of the vesicles and (g, h) phase diagrams for $A = 4.5$, $B = 2.02$, $\eta = 0.1$, and $D_u = 20$. (a, b, g) $V^* = 1$ (fixed shape). (c–f, h) $V^* = 0.8$. (a, c) $G\kappa_0/R^2 = 0.077$ and $C_0 R = 2$. (b) $G\kappa_0/R^2 = 0.061$ and $C_0 R = 8$. (d) $G\kappa_0/R^2 = 0$ and $C_0 R = 10$. (e) $G\kappa_0/R^2 = 0.046$ and $C_0 R = 8$. (f) $G\kappa_0/R^2 = 0.077$ and $C_0 R = 10$. The color in snapshots indicates the concentration of the curvature-inducing protein, $u$. Purple and green lines on the phase diagrams represent the Turing bifurcation curve and Hopf bifurcation, respectively, and the symbols represent the simulation results. The red line indicates $A + A' = 0$. Two or three overlapped symbols indicate the coexistence of multiple patterns.

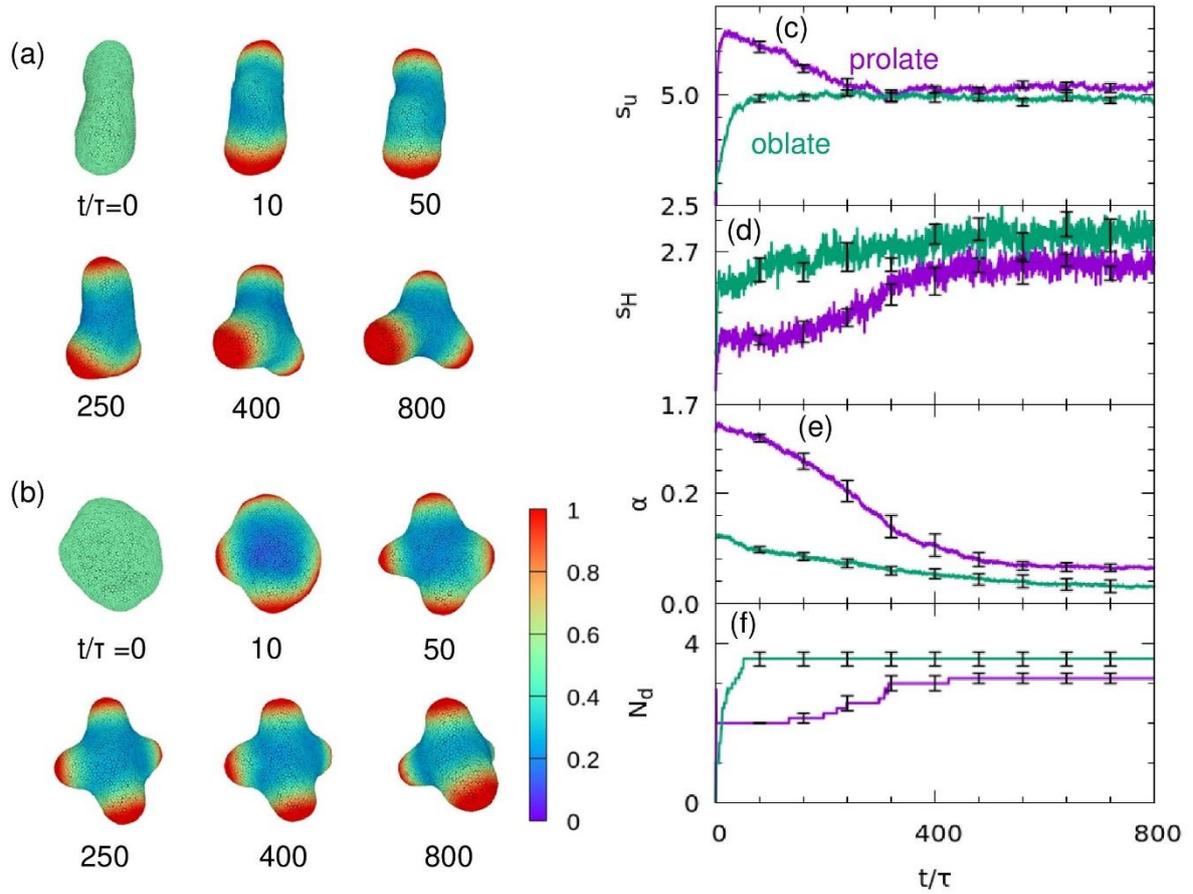

Figure 3. Examples of pattern formation and membrane deformation. (a, b) Sequential snapshots of the vesicles for $A = 4.5$, $B = 2.02$, $\eta = 0.1$, $D_u = 20$, $G\kappa_0/R^2 = 0.046$, $C_0 R = 8$, and $V^* = 0.8$ starting from (a) prolate and (b) oblate shapes. The color indicates the concentration of curvature-inducing protein, $u$. (c–e) Time evolution of (c) the separation metric of the protein concentration, $s_u$, (d) that of the local curvature, $s_H$, (e) asphericity, $\alpha$, and (f) the number of domains, $N_d$. The purple and green lines indicate the simulations starting from prolate and oblate shapes, respectively. Results are presented as the mean ± standard error ($n = 8$).

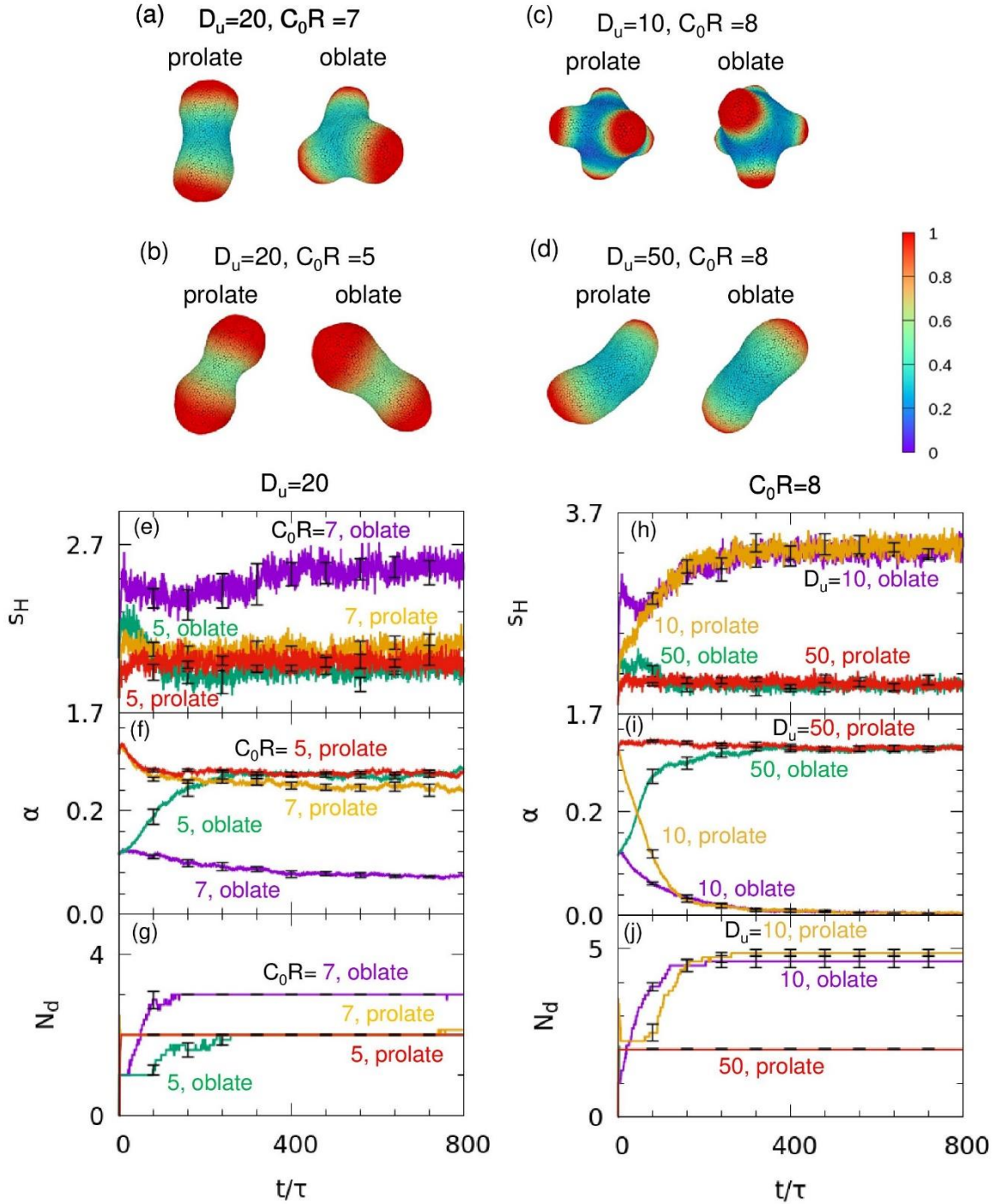

Figure 4. (a–d) Snapshots of vesicles for $A = 4.5$, $B = 2.02$, $\eta = 0.1$, $G\kappa_0/R^2 = 0.046$, and $V^* = 0.8$ for two values of $D_u$ and $C_0$ starting from prolate and oblate shapes. (a) $D_u = 20$ and $C_0R = 7$. (b) $D_u = 20$ and $C_0R = 5$. (c) $D_u = 10$ and $C_0R = 8$. (d) $D_u = 50$ and $C_0R = 8$. The color indicates the concentration of curvature-inducing protein, $u$. (e–j) Time evolution of (e, h) the separation metric of the local curvature, $s_H$, (f and i) asphericity, $\alpha$, and (g, j) the number of domains, $N_d$. The data for $C_0R = 7$ and $5$ at $D_u = 20$ are shown in (e–g), and the data for $D_u = 10$ and $50$ at $C_0R = 8$ are shown in (h–j). The orange

and red lines indicate simulations starting from prolate shapes, and the purple and green lines indicate simulations starting from oblate shapes. Results are presented as mean ± standard error ($n = 8$).

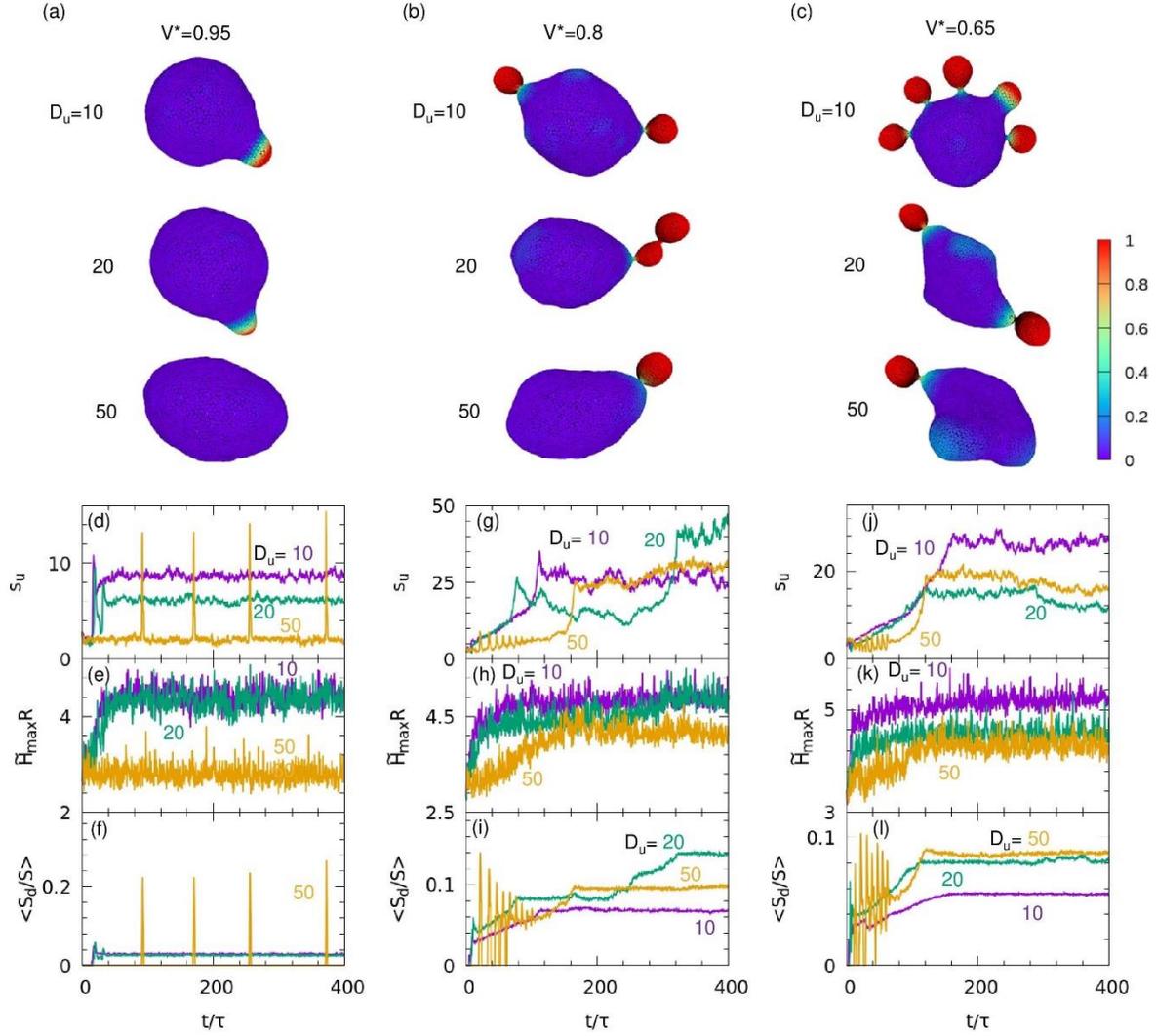

Figure 5. (a–c) Snapshots of the vesicles for $A = 4.5$, $B = 2.02$, $\eta = 0.1$, $G\kappa_0/R^2 = 0.077$, and $C_0 R = 10$ for three values of $V^*$ and $D_u$. (a) $V^* = 0.95$. (b) $V^* = 0.8$. (c) $V^* = 0.65$. The color indicates the concentration of curvature-inducing protein, $u$. (d–l) Time development of (d, g, j) the separation metric of the protein concentration, $s_u$, (e, h, k) the maximum value of the local curvature, $\widetilde{H}_{\max}R$, and (f, i, l) the mean area ratio of one domain $\langle S_d/S \rangle$. The data for $V^* = 0.95$, $0.8$, and $0.65$ are shown in (d–f), (g–i), and (j–l), respectively. The purple, green, and orange lines indicate the simulation data for $D_u = 10$, $20$, and $50$, respectively. Results of one typical simulation run are shown. The results averaged from eight independent simulations are shown in Supplementary Figure S4.

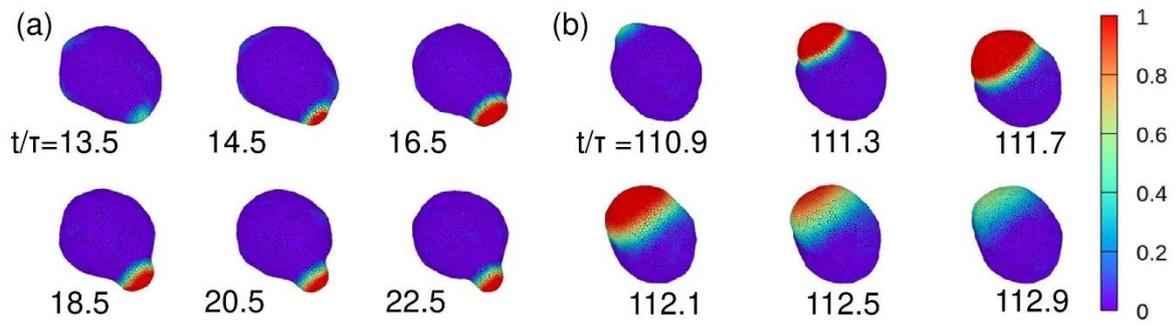

Figure 6. Sequential snapshots of the vesicles for $A = 4.5$, $B = 2.02$, $\eta = 0.1$, $G\kappa_0/R^2 = 0.077$, $C_0 R = 10$, and $V^* = 0.95$ for (a) $D_u = 10$ and (b) $D_u = 50$.

# Supplementary Material: Pattern Formation in Reaction-Diffusion System on Membrane with Mechanochemical Feedback

*Naoki Tamemoto and Hiroshi Noguchi**

Institute for Solid State Physics, University of Tokyo, Kashiwa, Chiba 277-8581, Japan

**CALCULATION OF DOMAIN SIZE AND LOCALLY AVERAGED CURVATURE**

A Turing pattern is considered to be formed when the minimum and maximum values of $u$ are more than 0.2 apart. We designate one region as the domain for the calculation of the number of domains $N_\mathrm{d}$ and the size of one domain $S_\mathrm{d}$ if $u$ is greater than the mean value of the maximum and minimum values and is connected by the bond network.

When evaluating the local curvature, we calculated $\widetilde{H}$ by averaging the curvature $H$ of the adjacent nodes up to depth 2 to reduce temporal thermal fluctuations. Without smoothing (i.e., smoothing depth = 0), the probability distribution of the local curvature is broad, and thus the dependence of the maximum values of local curvature $\widetilde{H}_\mathrm{max}$ on the diffusion constants $D_u$ is unclear (Figs. S1(b) and (c)). According to the probability distributions and the time evolutions of $\widetilde{H}_\mathrm{max}$, the shape differences can be distinguished well at the smoothing depth 2 or above (Figs. S1(b) and (c)). Thus, we chose the smoothing depth 2 because the local curvature can be evaluated and the summed area for smoothing is small (Fig. S1(a)).

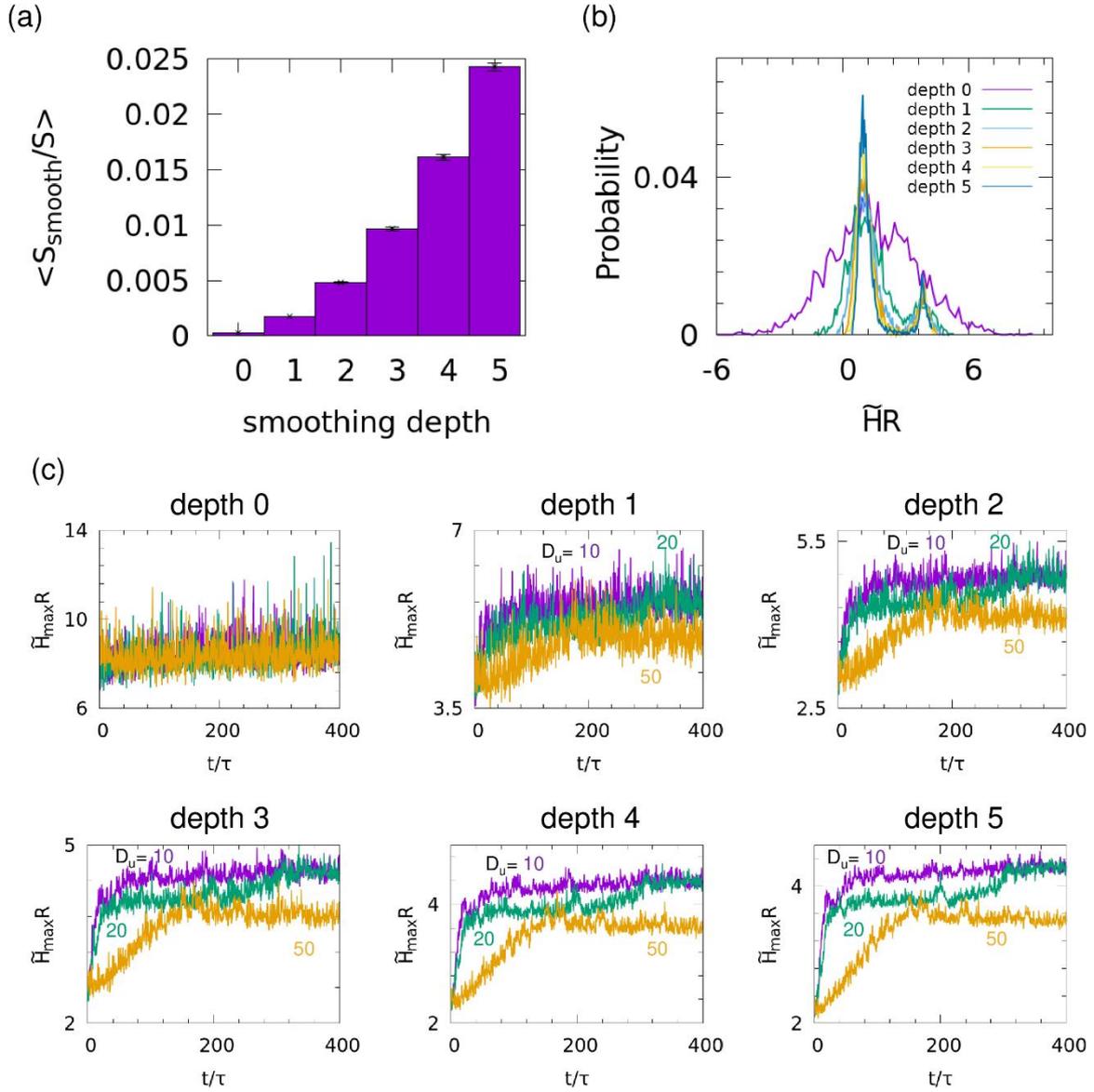

Figure S1. Dependence on smoothing depth to calculate the local curvature $\tilde{H}$ for the data shown in Fig. 5(h). (a) The average of a summed-area smoothed at various smoothing depths, $S_{\text{smooth}}$ as a proportion of total area $S$. (b) Probability distributions of the local curvature at various smoothing depths. The condition is the same as in (a). (c) Time evolutions of the maximum values of local curvature at various smoothing depths.

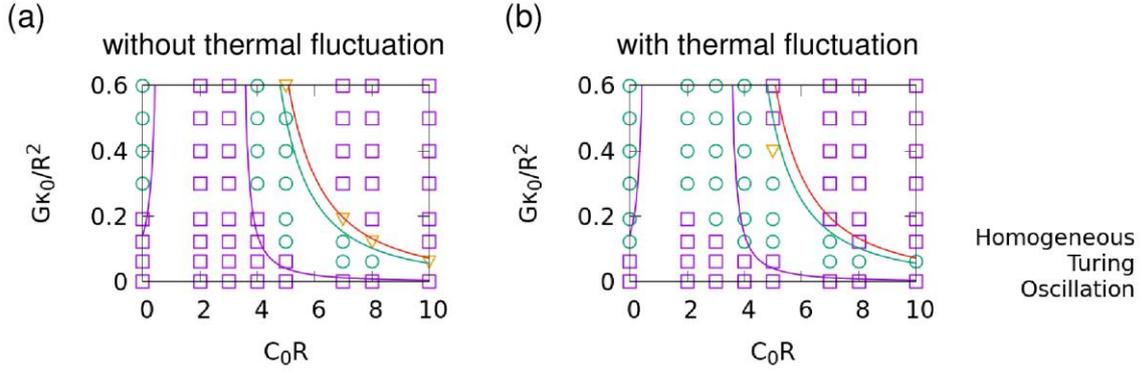

Figure S2. (a, b) Phase diagrams for $A = 4.5$, $B = 2.02$, $\eta = 0.1$, $D_u = 20$, and $V^* = 1$ without thermal fluctuations (a) and with thermal fluctuations (b). The purple and green lines on the phase diagrams represent the Turing bifurcation curve and the Hopf bifurcation, respectively. The symbols represent the simulation results. The red line indicates $A + A' = 0$.

**THE EFFECT OF THERMAL FLUCTUATIONS**

We performed simulations with and without thermal fluctuations at $A = 4.5$, $B = 2.02$, $\eta = 0.1$, $D_u = 20$, and $V^* = 1$ to investigate the effect of temporal thermal fluctuations (Fig. S2). Since the area and volume of the vesicles are constrained by the harmonic potentials, the reduced volume fluctuates only slightly, with a standard deviation of $\Delta V^* = 0.00004$. The results of the simulation without thermal fluctuations are consistent with those of the linear stability analysis. In contrast, for simulations conducted with thermal fluctuations, the results showing large $G$ are different from those of both the linear stability analysis and the simulation without thermal fluctuations (Figs. S2(a) and (b)). Thus, the stable phase is modified even by small membrane fluctuations.

This fluctuation effect is caused by the dependence of $A'$ on $H$. Since $A' = -G\left((\kappa_1 - \kappa_0)(2H)^2/2 - 2\kappa_1 C_0 H + \kappa_1 C_0^2/2\right)$, the average value of $A + A'$ in the simulation with thermal fluctuations is reduced by the variance of local curvature $H$. For example, in the simulation for $G\kappa_0/R^2 = 0.60$ and $C_0 R = 2$, the time-averaged values $\pm$ standard error of $A + A'$ are $3.83 \pm 0.01$ and $5.69$ in the presence and absence of thermal fluctuations, respectively. The latter value is in agreement with the theoretical value of $5.69$. The shift caused by the thermal fluctuations can be understood by the variance of the local curvature. When we eliminate the variance effect, the difference is removed as $A + A' + G(\kappa_1 - \kappa_0)\Delta(2H)^2/2 = 5.69 \pm 0.0004$, where $\Delta(2H)^2$ is the variance of the local curvature. Since the critical value of $A + A'$ for Turing bifurcation is $4.21$, the Turing patterns appear in the presence of thermal fluctuations. This effect becomes larger as $G$ increases, and thus the phase diagram with thermal fluctuations shows larger deviations at a larger value of $G$.

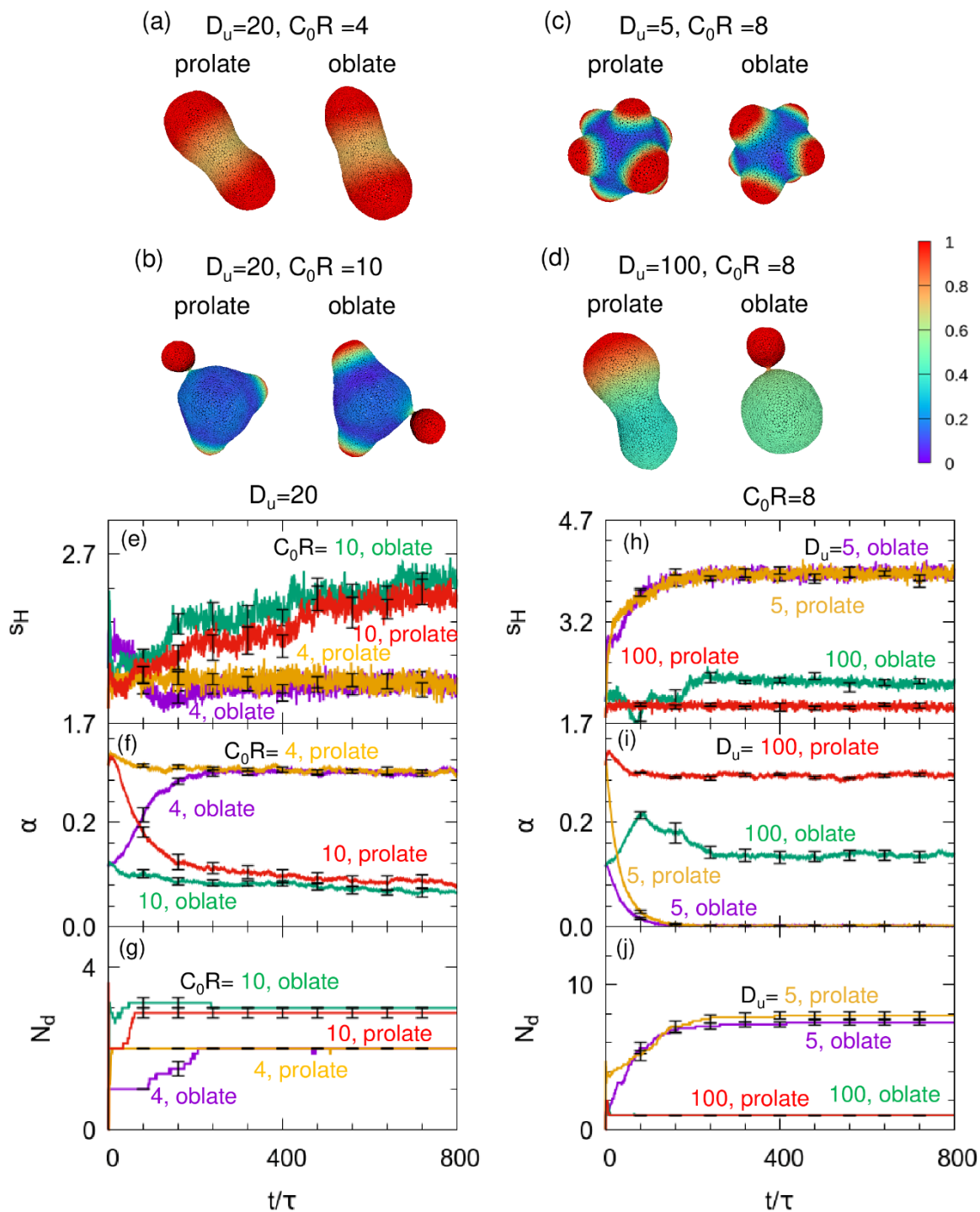

Figure S3. (a–d) Snapshots of vesicles for $A = 4.5$, $B = 2.02$, $\eta = 0.1$, $G\kappa_0/R^2 = 0.046$, and $V^* = 0.8$ for two more values of $D_u$ and $C_0$ starting from prolate and oblate shapes. (a) $D_u = 20$ and $C_0R = 4$. (b) $D_u = 20$ and $C_0R = 10$. (c) $D_u = 5$ and $C_0R = 8$. (d) $D_u = 100$ and $C_0R = 8$. The color indicates the concentration of curvature-inducing protein, $u$. (e–

j) Time evolution of (e, h) the separation metric of the local curvature, $s_H$, (f and i) asphericity, $\alpha$, and (g, j) the number of domains, $N_d$. The data for $C_0R = 4$ and $10$ at $D_u = 20$ are shown in (e–g), and the data for $D_u = 5$ and $100$ at $C_0R = 8$ are shown in (h–j). The orange and red lines indicate simulations starting from prolate shapes, and the purple and green lines indicate simulations starting from oblate shapes. Results are presented as mean ± standard error ($n = 7$ for $D_u = 100$ and $C_0R = 8$ started from oblate form. Otherwise $n = 8$).

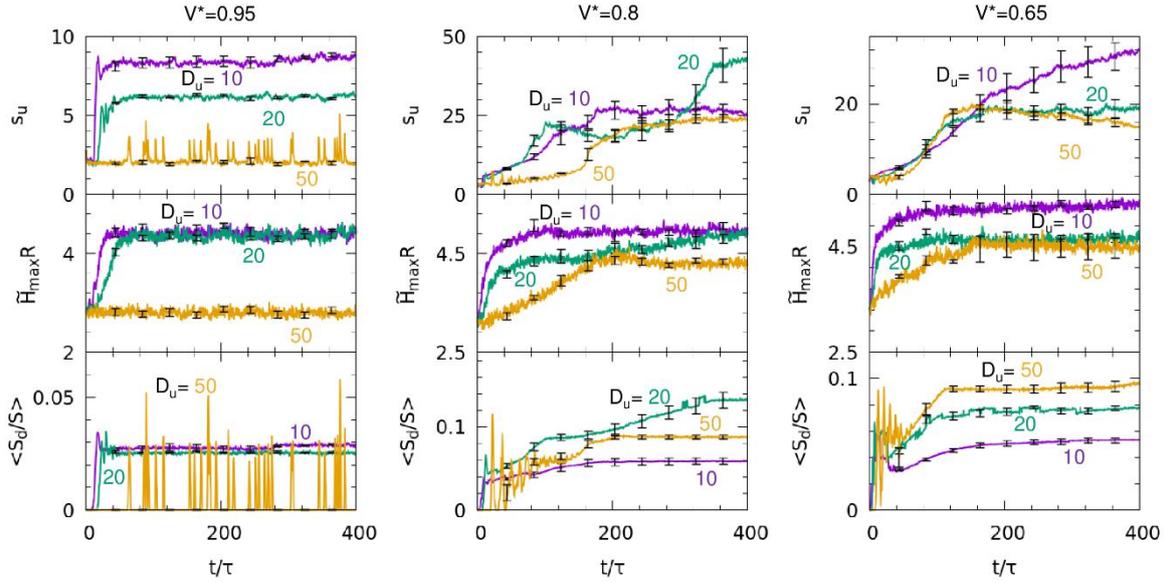

Figure S4. Average time evolution of the separation metrics, $s_u$, the maximum value of the local curvature, $\tilde{H}$, and the domain area ratio $\langle S_d/S \rangle$. Results are presented as the mean ± standard error ($n = 8$). The parameters are the same as in Fig. 5.

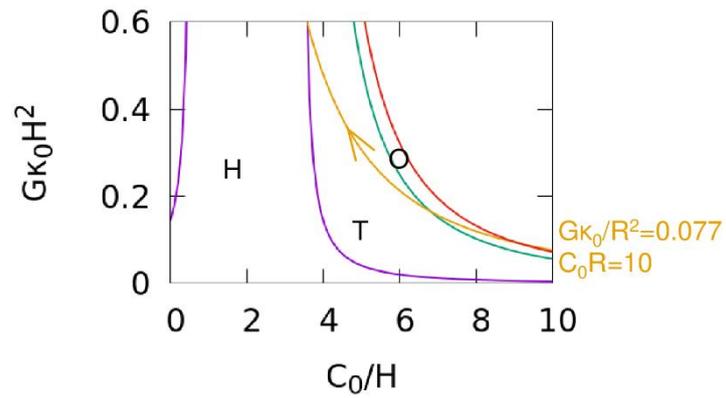

Figure S5. The phase diagram for the Brusselator, modified to include a membrane curvature effect, for $A = 4.5$, $B = 2.02$, and $\eta = 0.1$. The purple, green, and red lines are the same as those shown in Fig. 1. The orange line indicates stability in changing local curvature $H$ at $G\kappa_0/R^2 = 0.077$ and $C_0 R = 10$. As $H$ increases, it is shifted toward the upper left, and the transition from oscillation to Turing mode occurs.